\documentclass[journal]{IEEEtran}
\usepackage{cite}
\usepackage{amsmath,amssymb,amsfonts,amsthm}
\usepackage{graphicx}
\usepackage{epstopdf}
\usepackage{algorithm}
\usepackage{algorithmic}
\usepackage{enumerate}
\usepackage{subfigure}

\theoremstyle{plain}

\begin{document}

\title{Machine Learning Empowered Beam Management for Intelligent Reflecting Surface Assisted MmWave Networks}

\author{\IEEEauthorblockN{~Chenglu~Jia,~Hui~Gao,~Na~Chen,~and~Yuan~He}
\thanks{C. Jia, H. Gao, and Y. He are with the Key Lab of Trustworthy Distributed Computing and Service, Beijing University of Posts and Telecommunications, Beijing 100876, China, (e-mail: \{chenglujia, huigao, yuanhe\}@bupt.edu.cn); N. Chen is with Nara Institute of Science and Technology, Japan, (e-mail: \{chenna\}@is.naist.jp).}}

\maketitle

\begin{abstract}
Recently, intelligent reflecting surface (IRS) assisted mmWave networks are emerging, which bear the potential to address the blockage issue of the millimeter wave (mmWave) communication in a more cost-effective way. In particular, IRS is built by passive and programmable electromagnetic elements that can manipulate the mmWave propagation channel into a more favorable condition that is free of blockage via judicious joint BS-IRS transmission design. However, the coexistence of IRSs and mmWave BSs complicates the network architecture, and thus poses great challenges for efficient beam management (BM) that is one critical prerequisite for high performance mmWave networks. In this paper, we systematically evaluate the key issues and challenges of BM for IRS-assisted mmWave networks to bring insights into the future network design. Specifically, we carefully classify and discuss the extensibility and limitations of the existing BM of conventional mmWave towards the IRS-assisted new paradigm. Moreover, we propose a novel machine learning empowered BM framework for IRS-assisted networks with representative showcases, which processes environmental and mobility awareness to achieve highly efficient BM with significantly reduced system overhead. Finally, some interesting future directions are also suggested to inspire further researches.
\end{abstract}


%
\IEEEpeerreviewmaketitle

\section{Introduction}
Millimeter Wave (mmWave) band operating at 30-300GHz offers a large amount of unoccupied spectrum resource for ever-growing data-rate requirements \cite{6515173}. However, mmWave communications suffers from significant path loss and is highly sensitive to channel variation and blockage. To address these challenges, directional narrow-beam transmission is required to provide sufficient link budget between transceivers by using sophisticated array techniques. The establishment of such directional links requires precise beam alignment by proper beam management (BM). Note that the BM process at the initial access stage is also almost blind, and may drastically delay the access procedure. Up to date, the BM of mmWave has attracted much attentions from both industry, such as the 3rd Generation Partnership Project (3GPP), and academia \cite{giordani2018tutorial}.

The high directionality coupled with poor penetrability makes mmWave transmission vulnerable to blockage, which significantly degrades the coverage capability and limits the applications of mmWave in mobile cellular systems. Recently, intelligent reflecting surface (IRS), which is capable of achieving smart radio environment by manipulating the wireless propagation channel in a programmable manner, has been proposed to improve the spectrum and energy efficiency of wireless communication systems mostly working at the sub-6GHz band \cite{8647620, huang2019holographic, huang2019reconfigurable}. Specifically, IRS is a metasurface that is composed of a large number of Positive-Intrinsic-Negative (PIN) diodes. Each PIN diode is connected to a low-cost passive reflecting element, whose phase and amplitude can be controlled according to specific wireless radio environment and the target channel utility \cite{8647620}. Most recently, IRS is also applied to the mmWave networks to cost-efficiently address the blockage issue by creating the IRS-aided \emph{virtual} light-of-sight (LoS) transmission \cite{wang2019intelligent}, which avoids the densely deployment of mmWave BSs at very high costs. In particular, the joint beamforming designs have been focused by currently literatures, which verify that the presence of IRS can improve the mmWave coverage obviously. Those observations indicate that IRS can provide an energy-efficient and low-cost solution to address the blockage issue of mmWave networks \cite{7091932}. However, the presence of IRS complicates the original BS-UE one-hop communication topology into the BS-IRS-UE two-hop communication topology, which in turn calls for novel transmission protocols and schemes to make best use of IRS for performance enhancement. Moreover, most of the existing works assume prefect BM for simplicity, neglecting practical aspects and influence of BM in the IRS-assisted mmWave networks. To this end, it is imperative to design and evaluate the efficient BM mechanism for IRS assisted mmWave networks.

 The BM for conventional one-hop mmWave links is often achieved by time-consuming beam-space searching, which is not suitable for the more complicated IRS-assisted two-hop mmWave links \cite{giordani2018tutorial}. Conventional BM scheme often consists of two procedures, \emph{beam training} for the idle users and \emph{beam tracking} for the connected users. More specifically, the purpose of beam training is to achieve initial access for the idle users or obtain the initial channel state information (CSI), i.e., establishing an initial link from BS to UEs. Once the users are connected to the network with initial CSI, the beam tracking procedure is conducted to maintain the mmWave links, i.e., achieving fast beam alignment based on the prior link information. It is noted that, the BS-IRS-UE two-hop communication topology complicates the BM process, and prohibitive overhead or poor performance would be generated if the BM mechanisms for conventional one-hop mmWave links are straightforwardly applied to the IRS-assisted networks. Therefore, novel and specialized BM mechanisms are required to facilitate efficient link establishing and maintenance for the new paradigm, especially using some state-of-the-art techniques.
\begin{figure}[tp]
  \centering
  \includegraphics[width=0.45\textwidth]{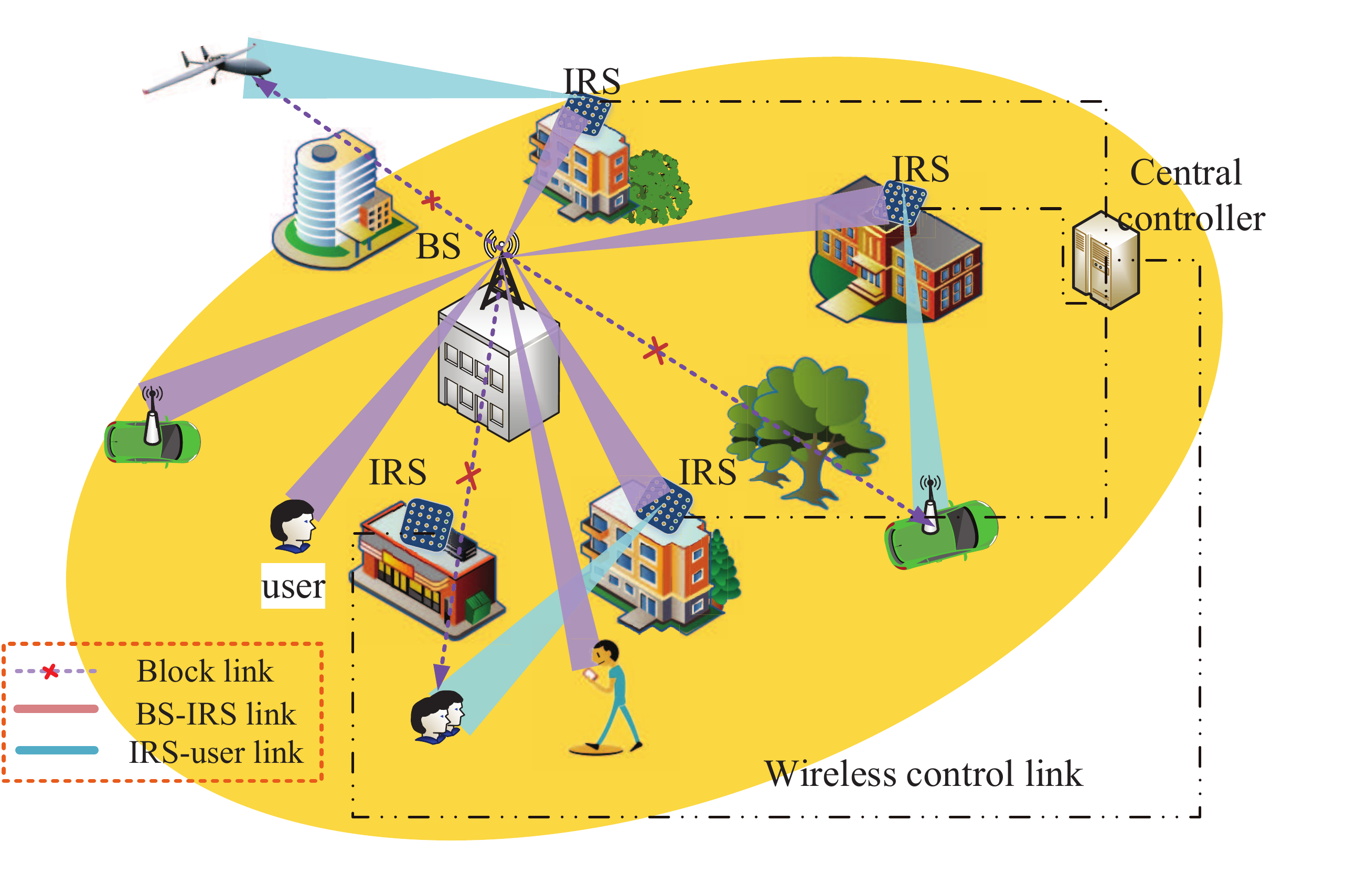}
  \caption{Illustration of the general IRS assisted mmWave network in urban areas. There are multiple IRSs and multiple pieces of UE, including cellphone, vehicles, unmanned aerial vehicles and etc, in the network. There are two access modes for those devices. On one hand, these devices can communicate with BS directly if there are reliable light-of-sight (LoS) links. Otherwise, when the BS-UE LoS links are blocked, reliable communications can be achieved with the assistance of IRSs.}
  \label{scenario}
\end{figure}

Recently, machine learning (ML) has been widely applied into the design and optimization of wireless communication systems for its powerful capabilities to tackle the complicated problems that are difficult to describe by a specific mathematical models or solved efficiently with conventional optimization tools. Notably, the applications of ML into mmWave networks are now emerging but are still very limited, especially for BM schemes. The BM for IRS assisted mmWave networks is a very challenging issue, which may require an even more complicated searching procedure to find a near-optimal solution in real time. In order to address this crucial challenge, one may expect ML as a potential and potent solution by learning from the previous BM, which is also the focus of our work.

\begin{table*}[htbp]
\centering
\caption{The Evaluation for The Existing BM Schemes}
\begin{tabular}{p{0.1\textwidth} p{0.1\textwidth} p{0.2\textwidth} p{0.4\textwidth}}
\hline
Reference & BM Mode & Method & The challenges when extend to IRS assisted mmWave network\\
\hline
\cite{6140087} & Beam training  & Conventional: narrow beam exhaustive searching  & High overhead\\
\hline
\cite{7885089} & Beam training & Conventional: wide beam hierarchical searching  & The passivity of IRS makes the frequent information exchange difficult\\
\hline
\cite{aviles2016position} & Beam training & Conventional: narrow beam searching with the aid of the user position  & Highly relying on the modeling of user's mobility\\
\hline
\cite{va2019online} & Beam training & Machine learning: selecting and refining the beam pairs with two-layer online learning framework  & The handover between BS and IRS\\
\hline
\cite{jayaprakasam2017robust} & Beam tracking & Conventional: extended Kalman filter  & Highly relying on the modeling of user's mobility\\
\hline
\cite{zhu2018high} & Beam tracking & Conventional: pairs of auxiliary beams  & Limited performance and high overhead\\
\hline
\cite{zhang2019position} & Beam tracking & Machine learning: position prediction  & The handover between BS and IRS\\
\hline
\end{tabular}
\label{reference}
\end{table*}

In this paper, we firstly discuss some key issues and challenges of BM for IRS assisted mmWave networks. Then, we systematically evaluate and analyze the state-of-the-art BM mechanisms and their adaptability for the IRS-assisted new paradigm. Motivated by these observations, a novel ML empowered BM framework for IRS-assisted mmWave networks with representative showcases is proposed, which processes environmental and mobility awareness to achieve highly efficient BM with significantly reduced system overhead. Finally, we highlight a few potential research directions for future exploration.


\section{Key Issues of BM for IRS-assisted MmWave Networks}

 In this section, we firstly analyze the key issues of BM for IRS assisted mmWave networks, which mainly lay in three aspects, namely the more complicated network communication topology, limited hardware capability and more complicated BM for mobility scenarios. Then, we summarize the typical BM mechanisms for conventional mmWave networks, and evaluate their adaptability and performance when extended to the IRS-assisted mmWave network, which offers valuable insights to guide the new design.

  We consider a typical IRS-assisted mmWave network, where multiple IRSs are deployed to provide seamless and stable service for the target areas where the coverage of the mmWave BS is blocked by the obstacles as shown in Fig. \ref{scenario}. Specifically, the IRS-assisted mmWave network can be regarded as a two-hop mmWave network, where BS performs beamforming to transmit the signal to the target IRS(s) firstly, and then the IRS carries out phase adjustment of reflecting elements to steer the incident signal towards the desired UE. Based on the above analysis, we summarize the key issues of BM for IRS assisted mmWave networks, which are elaborated as follows.

\subsection{The Two-hop Topology of IRS-assisted MmWave Network}
   Although IRS enhances the coverage performance of conventional BS-serve-only mmWave networks, the network communication topology is also complicated at the same time. The BM is already a very time-consuming and challengeable task for conventional one-hop mmWave networks, while the two-hop paradigm will further make the overhead of BM unacceptable for two-hop beam alignment. Specifically, in order to achieve initial access for the idle users or acquire initial CSI, firstly, an initial searching procedure is required to determine the access mode of UE connected directly or indirectly with the assistance of IRS to the mmWave BS. Furthermore, the specific access point of the target UE is identified. To do this, a refined beam training procedure is performed to achieve precise beam alignment. It is noted that, to maintain the links for the connected users, prior link information obtained from previous BM process, which consists of the position information, access mode, CSI and etc., can be fully utilized to achieve fast beam tracking. Here, we only provide a sketchy tentative plan for BM, while the practical implementation needs further study in detail. Therefore, more efficient and agile BM mechanisms dedicated to the complicated two-hop communication topology, are required to tackle two-hop beam alignment for both BS-IRS and IRS-UE links.

\subsection{The Passivity of IRS}
  In general, the passive IRS does not possess any signal processing/receiving/transmitting capabilities, which poses great challenges for BM. On one hand, the passivity makes it difficult to estimate CSI of the two-hop BS-IRS-UE channel, specifically for directional mmWave channel. Although the authors in \cite{taha2019enabling} proposed that IRS equips the active radio frequency chains to estimate BS-IRS and UE-IRS channels, respectively, while it contradicts the initial intention to utilize the passive IRS for energy-efficient mmWave networks. In particular, the angular domain information (ADI), a critical component of CSI, can be the key prior information for BM. On the other hand, the implementation of BM requires frequent information exchange among BS, IRS and users, while passivity of IRS makes the information exchange among BS, IRS and users more difficult. Therefore, it is necessary to design the efficient ADI acquirement and information exchange mechanisms to overcome the challenges induced by the passivity of IRS.


\subsection{Handover for Multi-point Beam-switching}

  The existing BM mechanisms for conventional one-hop mmWave networks mainly focus on the quasi-static scenarios where the low-speed users are moving within the limited areas, while there is few discussion about handover for the high-mobility applications, such as unmanned aerial vehicle (UAV). In such case, link interruption may frequently take place when mobile terminals move between different cells. When it comes to the IRS-assisted mmWave network, more frequent handover is required to guarantee link quality of the users in real time since the two-hop communication topology gives rise to more selections of multi-point beam-switching, i.e., handover between IRSs within the same cell or handover between different cells. In this way, the concept of handover is further expanded, and conventional handover mechanisms may be not suitable for the new network paradigm. Therefore, more agile handover mechanism for multi-point beam-switching is required to achieve better mobility support for IRS assisted mmWave networks.

  \begin{figure}[tp]
  \centering
  \includegraphics[width=0.48\textwidth]{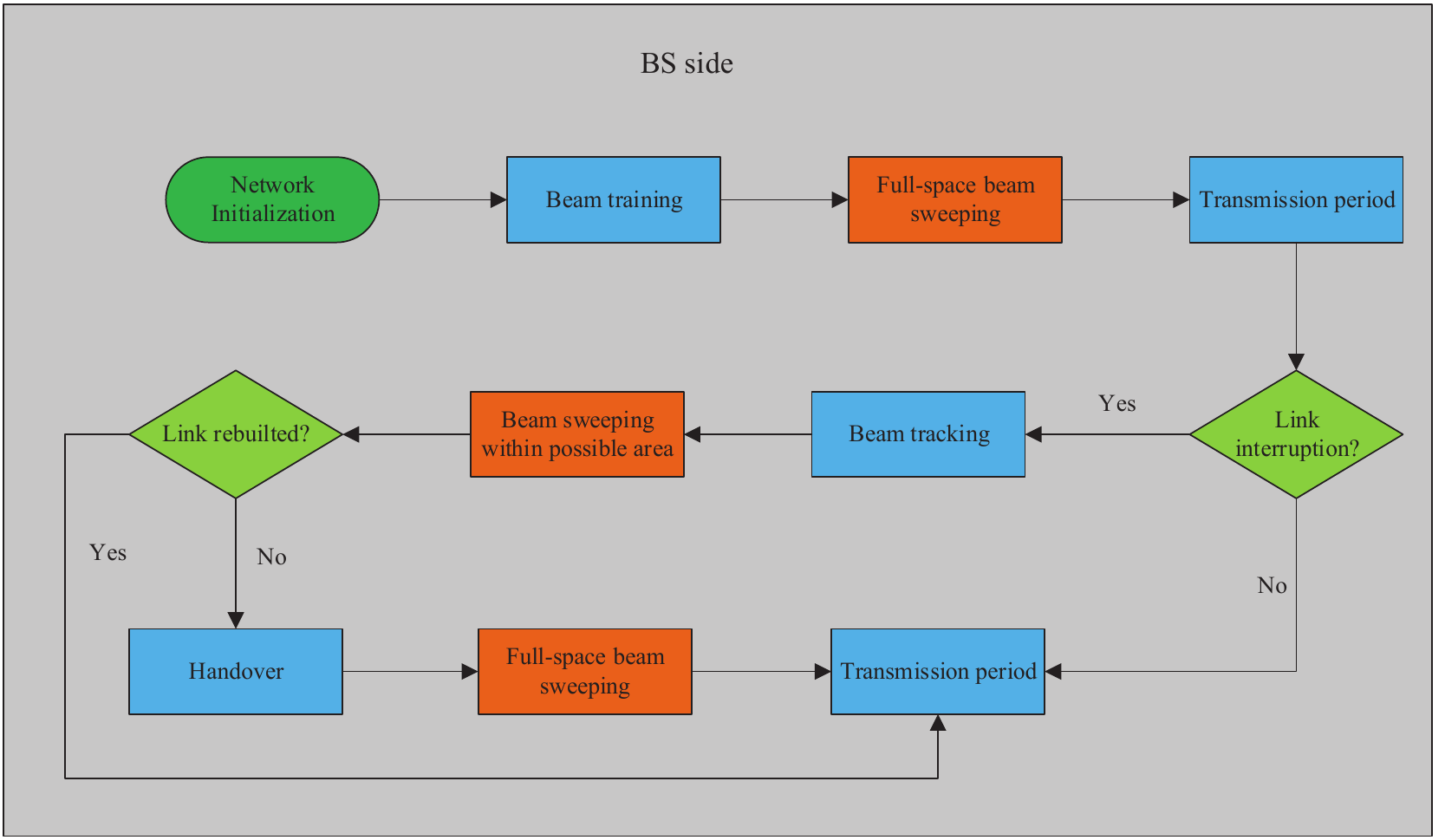}
  \caption{The general BM framework for IRS assisted mmWave networks.}
  \label{framework}
\end{figure}

\subsection{Extensibility of The Existing BM Mechanisms}
 In this subsection, we discuss the extensibility of existing BM schemes of conventional one-hop mmWave networks towards the IRS-assisted mmWave networks. For clarity, different BM mechanisms adopted at conventional mmWave networks are evaluated as Table \ref{reference}. Notably, the majority of these studies focus on beam-space searching based schemes. Specifically, to achieve initial access for the idle users, the authors in \cite{6140087} proposed a sequential beam training protocol, where BS performs narrow-beam exhaustive searching within the full space. However, the training overhead of the sequential beam training protocol is already prohibitive for the one-hop mmWave network, and the overhead will grow exponentially when extended to the IRS-assisted two-hop networks. Aiming to improve the efficiency of beam training, hierarchical beam training protocol is further designed by combining wide-beam and narrow-beam searching \cite{7885089}. Utilizing the position statistic of the user, the authors in \cite{8736395} present a fast beam training protocol by narrowing the searching range dynamically. It is noted that these ideas can be the essential guidance to the design of BM for IRS assisted mmWave networks. To maintain the mmWave links, extended Kalman filter \cite{jayaprakasam2017robust} is utilized for beam tracking, whose performance highly relies on the precise modeling of user motion. Then, the authors in \cite{zhu2018high} proposed an auxiliary beams pairs assisted beam tracking scheme, while high overhead and limited performance will be generated when extend to the IRS-assisted mmWave networks due to its awkwardness in handover. Recently, ML is emerging as a promising technology to achieve efficient BM for conventional mmWave networks. Specifically, the authors in \cite{va2019online, zhang2019position} proposed the ML based BM mechanisms by using the position information of users, which significantly improve the efficiency of BM compared to their counterparts. However, it is still impossible to fully address the challenges of BM for IRS assisted mmWave networks.

 Motivated by the above analysis, we can conclude that the BM for IRS assisted mmWave networks faces many challenges, such as high overhead, limited performance, frequent handover, and etc.. Most importantly, there are limited researches on BM issue for the new network topology, while efficient BM is the prerequisite to enable the high-performance mmWave network. As such, considering the complexity of BM issue for IRS assisted mmWave networks, we resort to ML to address these challenges efficiently.

\section{Machine Learning Empowered BM Framework}

 Motivated by the concept that the best BM mechanism relies on the specific environment \cite{giordani2018tutorial}, we combine the awareness capability of ML into BM to address the challenges for IRS assisted mmWave networks. In this section, a novel ML empowered BM framework is proposed to achieve agile BM for IRS assisted mmWave networks.

\subsection{General BM Framework}
\begin{figure}[tp]
  \centering
  \subfigure[]{\includegraphics[width=0.45\textwidth]{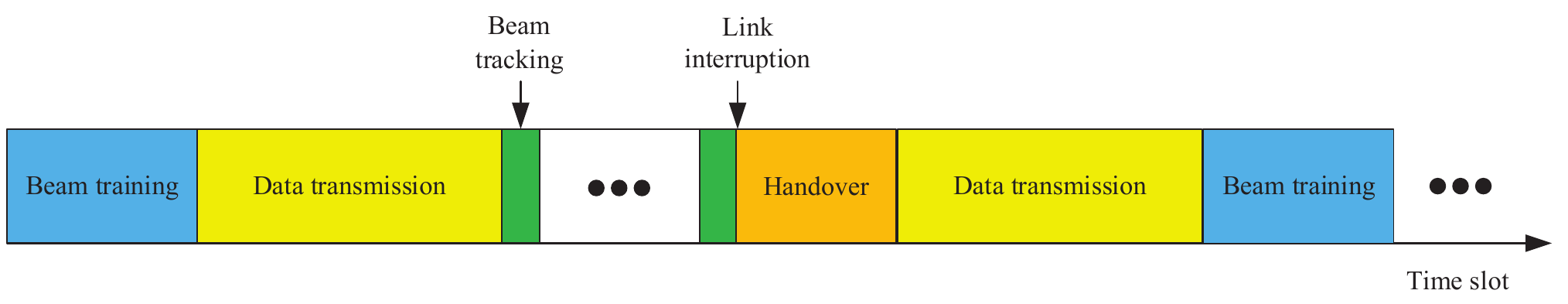}}
  \subfigure[]{\includegraphics[width=0.45\textwidth]{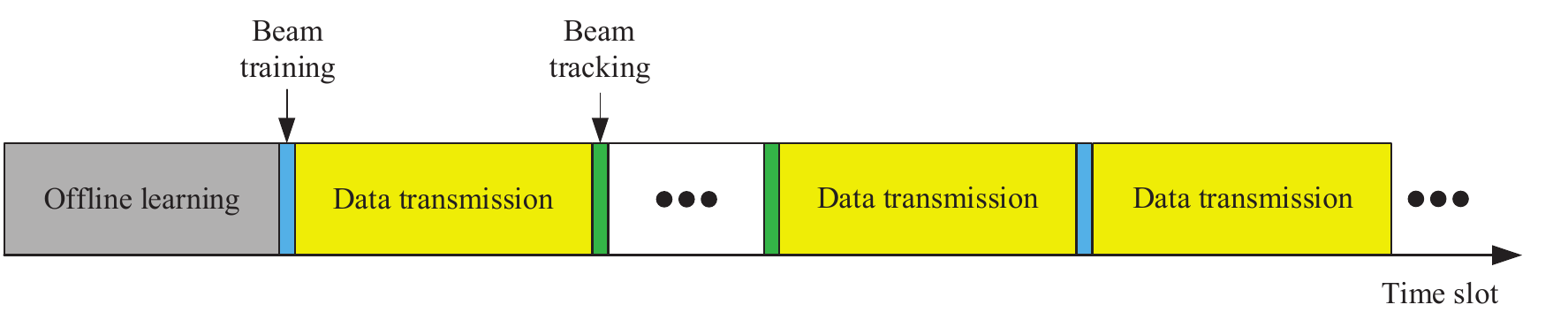}}
  \caption{Time-frame structure of IRS assisted mmWave networks: (a) conventional BM framework; (b) ML empowered BM framework.}
  \label{frame}
\end{figure}

In this part, by integrating ML tools, we propose a general BM framework for IRS assisted mmWave networks to show its potential, as shown in Fig. \ref{framework}. In general, BM often consists of the time-consuming beam-space searching approaches to achieve beam alignment. For illustration purpose, we first present a time-frame structure to clarify the whole transmission process, as shown in Fig. \ref{frame}, which consists of beam training for initial access and channel acquisition, data transmission, and beam tracking for link maintenance. As mentioned early, utilizing the awareness capability of ML, it is expected that the ML empowered BM may consume less time for beam alignment, and fasten handover or beam switching. Consequently, more resource can be reserved for data transmission.

 For general BM, firstly, the beam training procedure is performed to connect the idle users to the network, i.e., initial access. Then, the end-to-end communications among BS and users are conducted for both uplink and downlink. Sometimes, these directional links may be interrupted due to the mobility of users and blockages. Moreover, by utilizing the prior link information obtained from previous beam training/tracking, new beam tracking procedure is conducted to rebuilt the mmWave links quickly. However, beam misalignment may happen when the beam tracking procedure is applied to the high-mobility or complicated scenarios involving inter-BS or IRS beam switching. In such cases, an efficient handover or multi-point beam switching mechanism, acting as a necessary supplement for beam tracking, is required to guarantee link quality and seamless connections.

\begin{figure}[tp]
  \centering
  \includegraphics[width=0.45\textwidth]{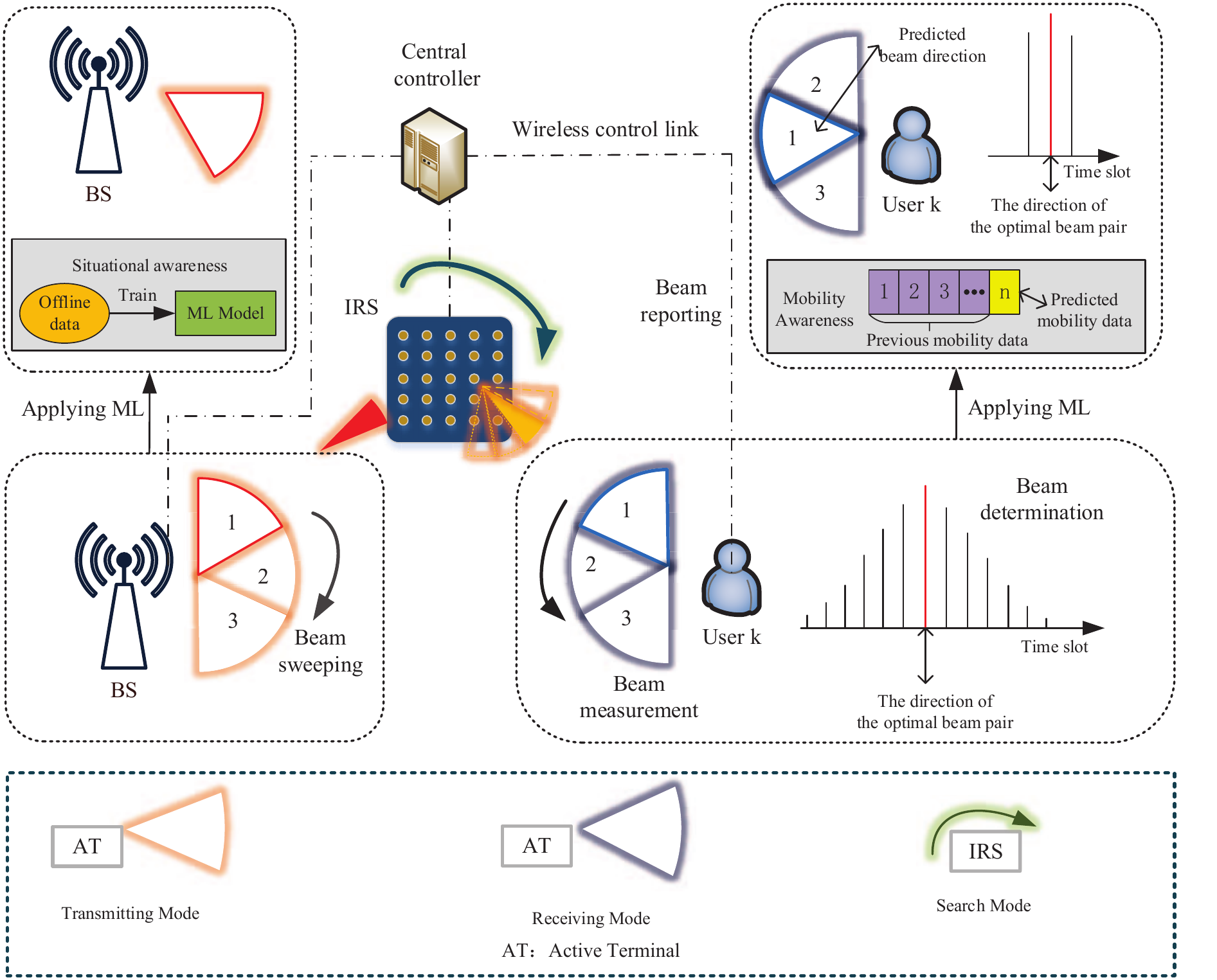}
  \caption{The beam-space searching based beam management.}
  \label{beam_training}
\end{figure}
The general BM framework highly depends on the beam-space searching based operations as shown in Fig. \ref{beam_training}, containing beam sweeping, beam measurement, beam determination, and beam reporting \cite{giordani2018tutorial}. Specifically, for the users connected to the mmWave BS directly, the mmWave BS transmits narrow beam carrying the training signals to sweep the full-space. At receivers, the multi-antenna users combine the signals sequentially, and then report the beam direction with maximal received power to the mmWave BS by wireless control links. While for the users connected to the mmWave with the aid of IRS, the BM process is more complicated. Firstly, the mmWave BS and IRS perform full-space exhaustive beam sweeping simultaneously, and the users combine the training signals sequentially until the maximal received power is obtained. Therefore, we can observe that the complicated network topology will further deteriorate the efficiency of BM. Fortunately, the popular ML tools can be integrated to acquire environmental awareness and mobility awareness, which bears the potential to enhance BM performance for IRS assisted mmWave networks. More details will be presented in next sub-section.

\subsection{ML Empowered BM Framework}

 \begin{figure}[tp]
  \centering
  \includegraphics[width=0.4\textwidth]{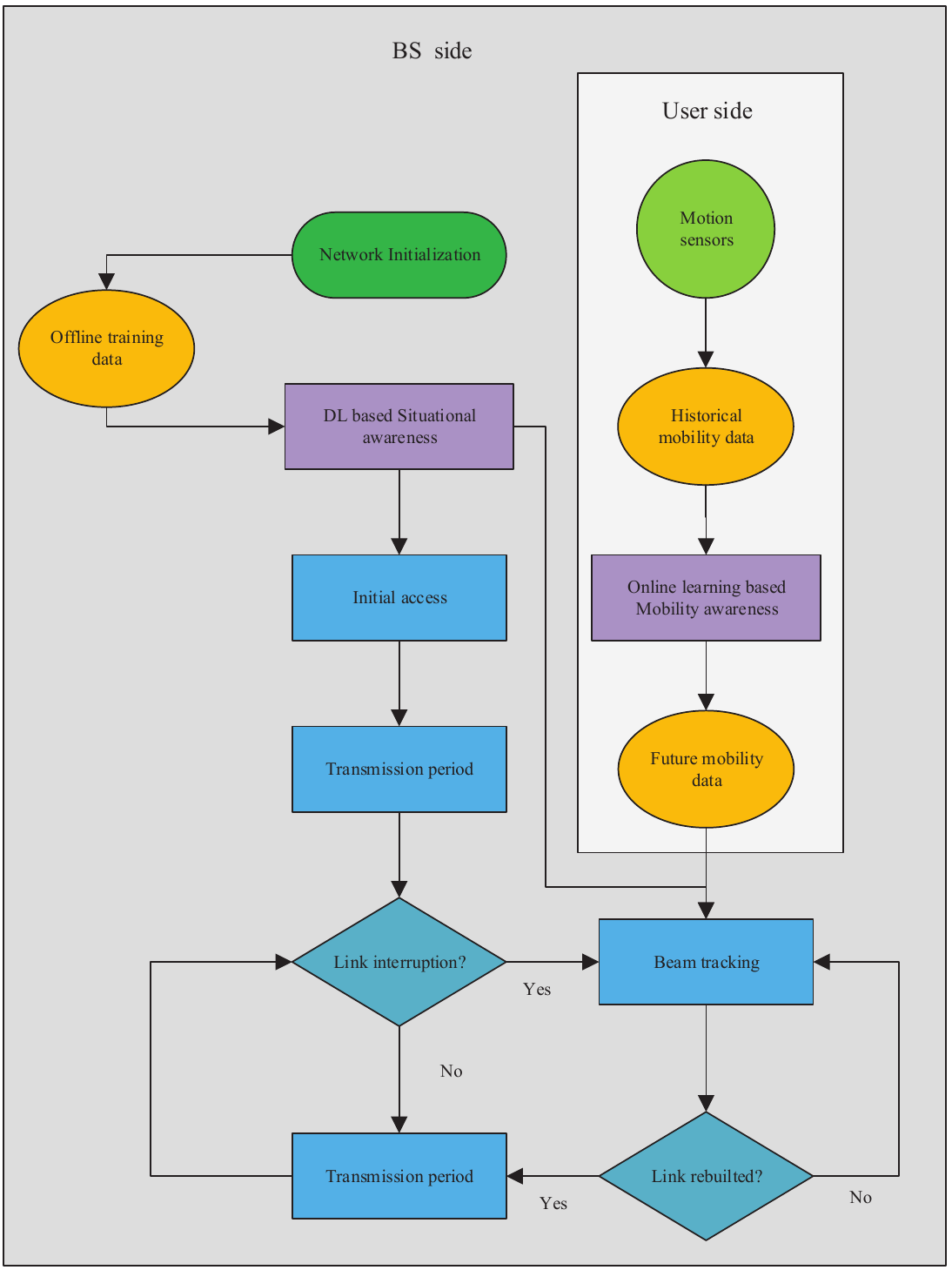}
  \caption{The ML empowered BM framework for IRS assisted mmWave networks.}
  \label{ML_framework}
\end{figure}
 The complicated two-hop network topology makes BM difficult to fit in a stochastic mathematical model, which motivates the exploration of data-driven BM. Given environment geometries, including the distribution of surrounding buildings and the deployment of IRSs, the mmWave channels are more deterministic compared to the sub-6GHz channel due to the highly directional transmission. Therefore, by utilizing the environmental awareness of ML, the BM information can be learned from previous BM with negligible performance loss. Meanwhile, mobile terminals, such as UAV, vehicle and etc., also travel under a certain mobility model, which makes their mobility information in a predictable way. Inspired by above observations, we propose a ML empowered BM framework as shown in Fig. \ref{ML_framework}. As compared to the conventional counterpart, the time-consuming beam sweeping procedure can be significantly simplified, which facilitates the efficient implementation of \emph{The Ultra Reliable Low Latency Communications} (uRLLC) for 5G and beyond.


\subsubsection{Deep Learning based Environmental Awareness}

At present, wireless communication devices are widely equipped with multiple sensors with different functions, such as Global Position System (GPS), radar and cameras, which can be utilized to localize the user and sense the environment \cite{gonzalez2017millimeter}. By fusing the information collected from these sensors, the position of users and the relevant environmental information can be acquired to make BS-IRS-UE \emph{connected} in real time. Based on the long-term statistics of the environment geometries, we resort to deep learning (DL) to encode the complicated environment geometries, also named environmental awareness. Specifically, DL, also named deep neural network (DNN) based ML, is a kind of supervised learning method, which regards the state of the network as input, such as CSI, position distribution of users, and etc.. Then, the optimal network parameters, such as beam direction, resource allocation and etc., are treated as label, so as to obtain a data-driven DNN model. In particular, the DNN can dig the complicated optimization relationship between real-time network state and optimal network performance. When the real-time network state is fed into the DNN model, the optimal network parameters can be predicted immediately without complicated optimization operations, which significantly degrades the system overhead. In this part, by regarding the position information of users as the network state, the efficiency of BM can be greatly improved for IRS assisted mmWave networks.
\begin{figure}[tp]
  \centering
  \includegraphics[width=0.45\textwidth]{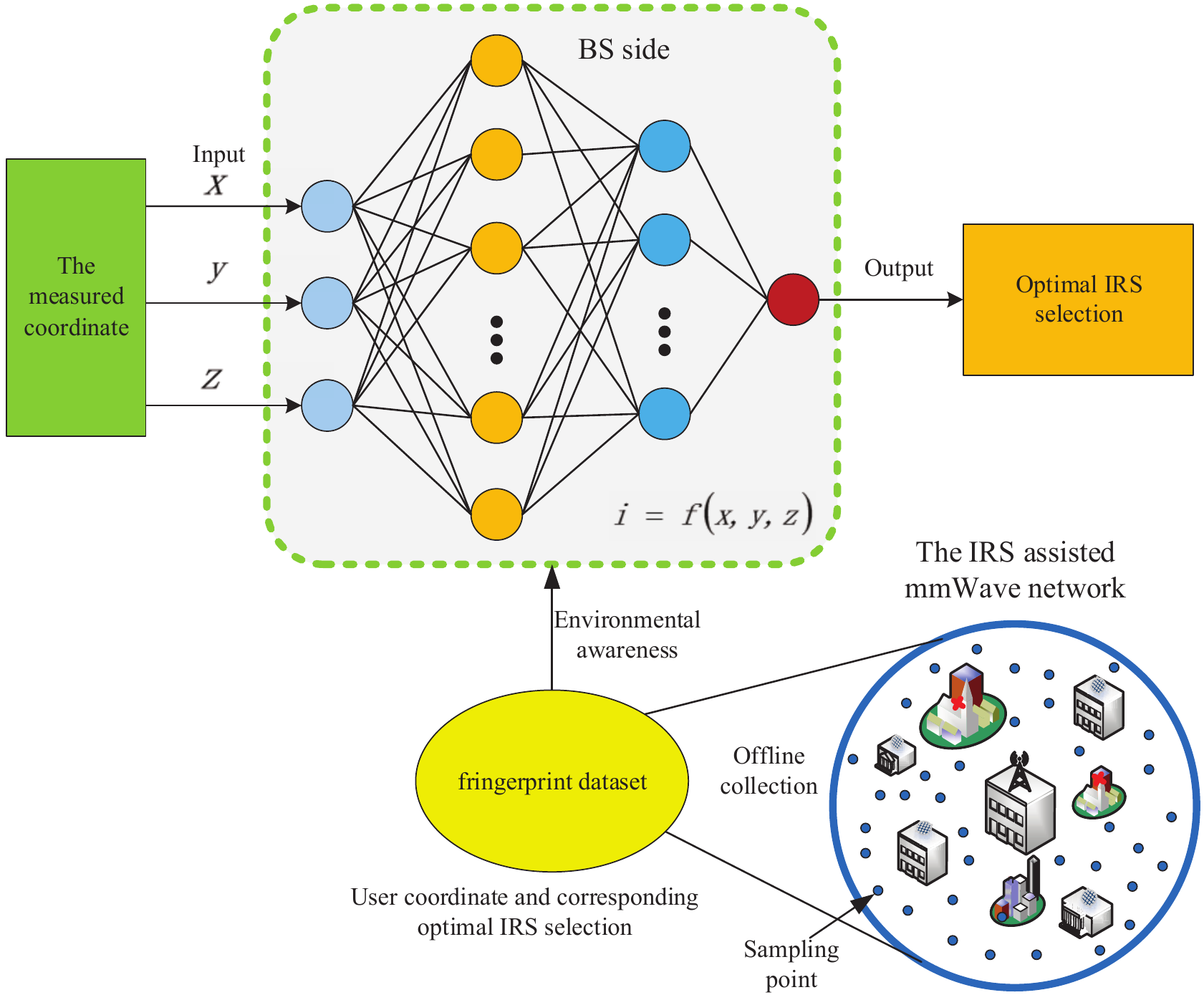}
  \caption{The deep neural network for environmental awareness.}
  \label{DL}
\end{figure}

Our method capitalizes on the fact that based on the cognition of environment, the corresponding BM decision can be identified according to the position information of users. One example of DNN is shown in Fig. \ref{DL}, wherein the DNN has been trained and learned offline with collected fingerprinting database so that IRS selection optimization can be sufficiently solved immediately. For example, when the three-dimension coordinate $\mathbf{p}=\left(x,y,z\right)$ of the target user is input, and the index of the optimal IRS, i.e., the best IRS and target user association that can generate best performance when the BS-UE direct link is blocked, can be immediately obtained at the output. Considering the specific implementation, when the users feed the position information back to the BS by broadcast channel, the BS can determine their optimal IRS selection. Then, the BS can directly steer the training beam to the corresponding IRS, and the IRS performs passive beamforming towards the corresponding position. Therefore, the frequent and time-consuming beam sweeping operation can be avoided by integrating ML into BM for IRS assisted mmWave networks when we compare the Fig. \ref{framework} and Fig. \ref{ML_framework}.

In order to achieve such favorable functionality with DNN, the fingerprinting database shall be established in advance. In particular, one one hand, the fingerprinting data can be collected by the sensors, and calculated by edge computing and stored by edge cache, and further the scalable, semi-distributed and efficient network platform can be established by integrating communications, sensing, localization, storage and computing with the aid of edge intelligence \cite{di2019smart}. On the other hand, big-data sharing is gradually recognized by both academia and industry, and more and more network databases can be acquired to support the application of artificial intelligence (AI) in future wireless networks \cite{xu2018data}. Therefore, with the increasing computing, storage and data acquisition capabilities, the communication oriented environmental awareness will make AI an indispensable tool to optimize and operate future wireless networks.


\subsubsection{Online Learning based Mobility Awareness}

The aforementioned environmental awareness based BM mainly captures the features of the quasi-static scenarios, which constitutes the static foundation of BM for IRS assisted mmWave networks. However, for mobility-enhanced mmWave applications, such as UAV, motion ground vehicle communications and etc., more features regarding the dynamic communication topology and its interplay with the static environment should be captured. More specifically, the high-mobility makes the channel change dramatically, and thus the more agile BM is required to adapt to the rapid channel or motion-resultant environmental variation. In this regard, the historical motion state information of the mobile terminals is exploited in conjunction with the environmental awareness, and the online learning methods dedicated to the prediction of future motion state information and short-term trajectory are applied to assist BM.

\begin{figure}[tp]
  \centering
  \includegraphics[width=0.45\textwidth]{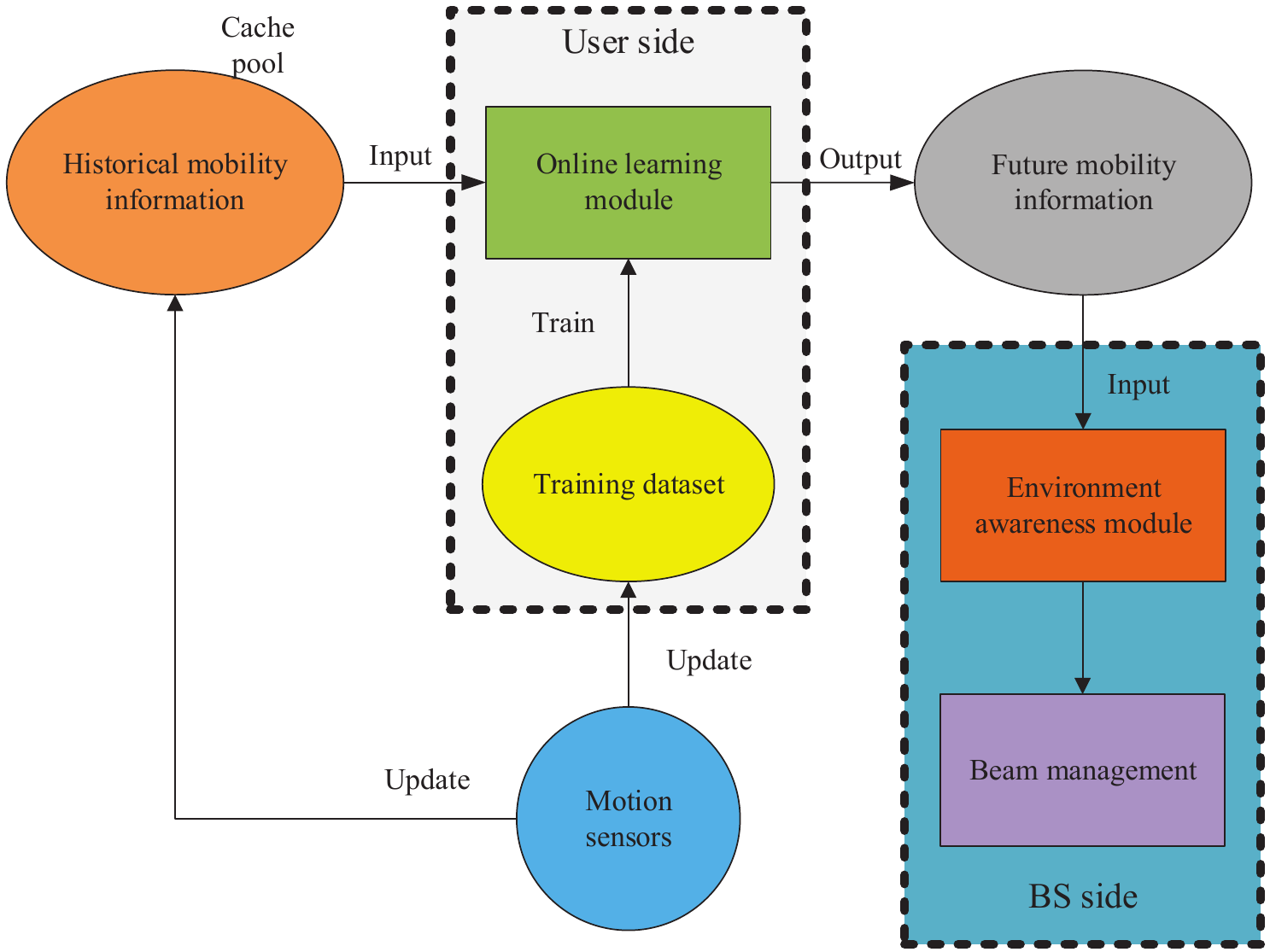}
  \caption{The online learning for mobility awareness.}
  \label{online_learning}
\end{figure}
At present, mobile terminals are often equipped with multiple motion sensors of different functions, such as GPS, Speed Sensor. The collected and fused mobility-related information can be utilized to capture the mobility features of the mobile terminals in real time, and then the data-driven ML model is established to predict the future motion state information based on the historical motion state information. In particular, the mobility awareness can be realized via the popular online learning methods, such as Gauss process, dynamic neural network and etc..

 The specific implementation of the online learning based mobility awareness is illustrated as Fig. \ref{online_learning}. There is a cache pool at the mobile terminal to store the mobility data captured by motion sensors. Based on the training dataset that are updated by motion sensors in real time, the online learning module is trained to capture the mobility law every several time slots. When historical motion state information is input into the online learning module, the mobility state information of future several time slots will be predicted immediately, and then when the predicted mobility information is input into the DNN, the optimal IRS selection can be further predicted. By integrating environmental and mobility awareness, on one hand, the proactive multi-point beam-switching for high-mobility devices, such as UAV, can be achieved by mapping the predicted position to the optimal IRS selection with the designed DNN, which avoids the prohibitive system overhead induced by the frequent handover. In this way, the mobile terminals can switch connected access point among IRSs quickly in response to the rapid-varying link quality, and the handover issue caused by high-mobility can be tackled efficiently. On the other hand, with reference to the predicted position and the corresponding IRS allocation, the beam tracking procedure can be greatly accelerated by searching the predicted area, which avoids frequent beam measurement operation. Therefore, by utilizing mobility awareness, the efficiency of BM can be enhanced for high-mobility scenarios.

\textbf{Summary}: The deployment of IRS significantly improves the flexibility and coverage of conventional mmWave networks, and meanwhile poses the new challenges for BM. In this subsection, the DNN is utilized to recognize the complicated network environment geometries, including the distribution of surrounding buildings and the deployment of IRSs. Moreover, the online learning method is adopted to predict the motion state information of the mobile terminals. In this way, by integrating environmental and mobility awareness, the aforementioned challenges of BM caused by the frequent handover and complicated network topology can be addressed efficiently for IRS assisted mmWave networks. In particular, the CSI is not necessary for proposed ML empowered BM framework. As such, the passivity of IRS will not give rise to the specific challenges for BM. Therefore, we can conclude that the ML empowered BM framework can address the aforementioned challenges efficiently and will play an essential role for the implementation of IRS-assisted mmWave networks.

\section{Study Case}

\begin{figure}[tp]
  \centering
  \includegraphics[width=0.45\textwidth]{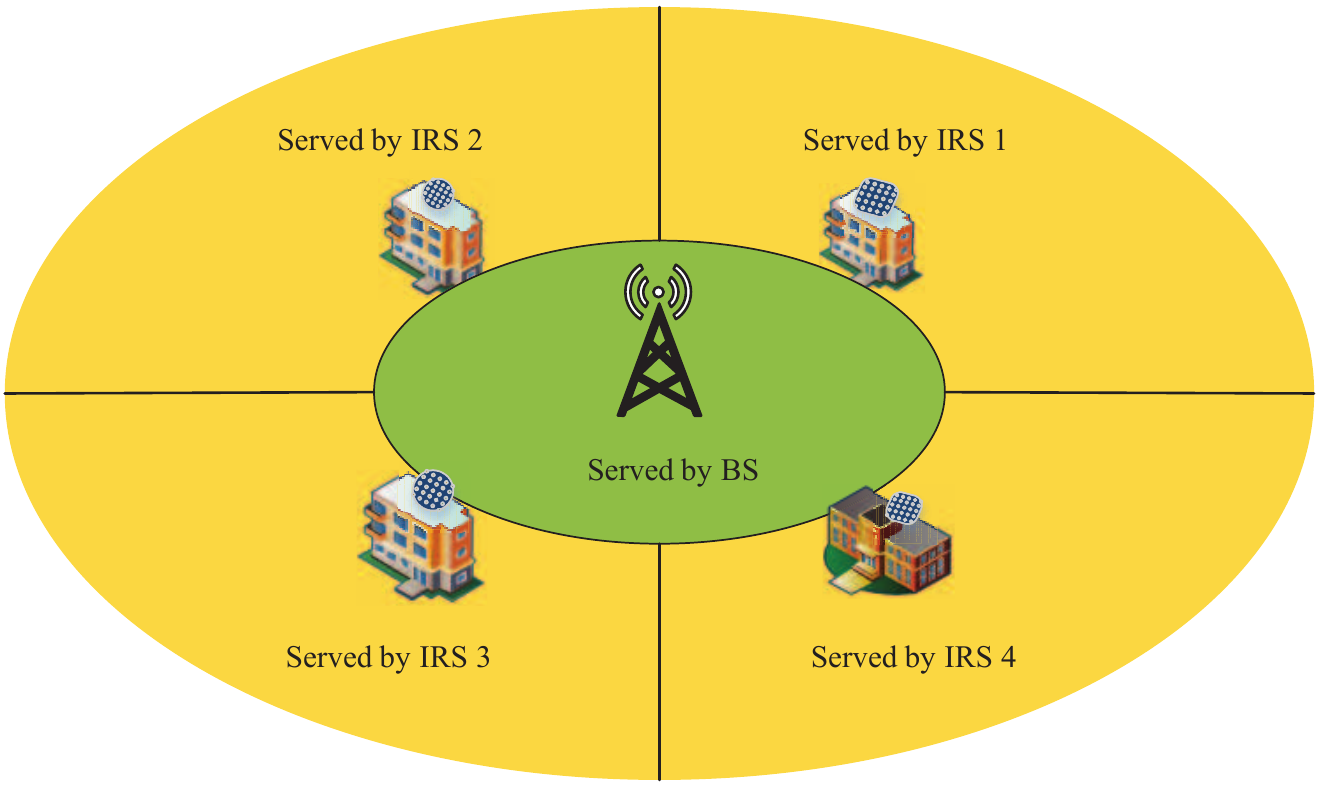}
  \caption{The specific study case for the IRS assisted mmWave network. Different colors represent the areas are served by different IRS or by BS.}
  \label{study_case}
\end{figure}

A specific study case is presented to verify the proposed ML empowered BM framework. For convenience of elaboration, we consider a simple yet typical IRS assisted mmWave network as shown in Fig. \ref{study_case}, where four IRSs are deployed to serve the areas that are not covered by the single BS due to blockages. Moreover, the network is divided into a large number of grids to deploy sampling points to create the fingerprint database by an offline way.

\begin{figure}[tp]
  \centering
  \includegraphics[width=0.45\textwidth]{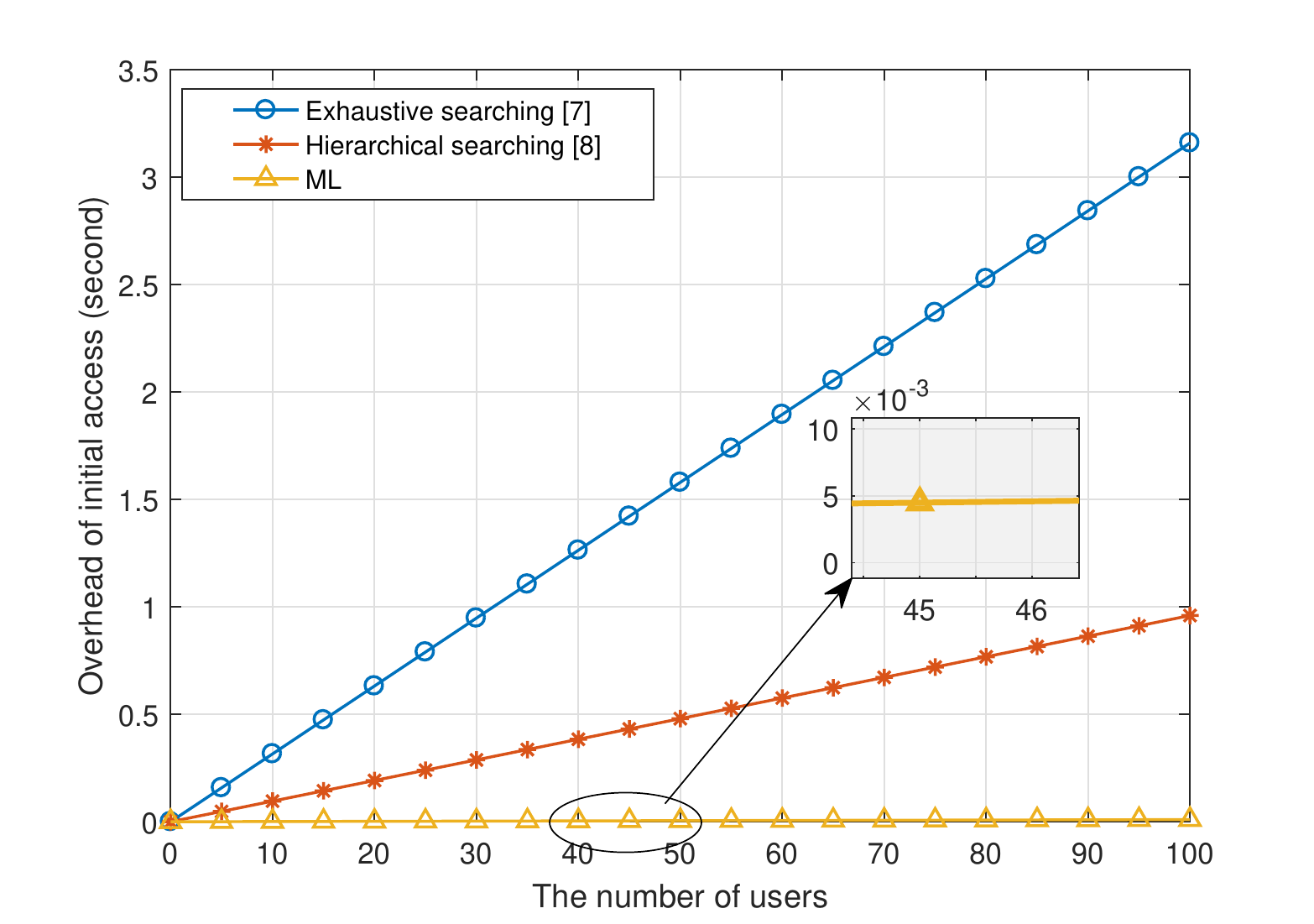}
  \caption{The comparison of different BM schemes over the overhead of initial access.}
  \label{overhead}
\end{figure}

In essence, the environmental awareness is a multi-classification problem with reference to the coordinate of online devices, which can be handled by  DNN. As shown in Fig. \ref{overhead}, we compare the system overhead of the proposed ML empowered BM with conventional exhaustive and hierarchical searching counterparts. Assuming that the duration of a time slot is 100 $\mu s$, we can observe that the system overhead of initial access increases linearly with the number of users, and obviously, the ML based BM mechanism is significantly superior to the exhaustive searching and exhaustive searching based schemes that suffer from unacceptable training overhead. Specifically, when the number of users reaches 100, the system overhead of initial access is about 3.1s for exhaustive searching, while the system overhead of ML empowered BM is negligible. Therefore, the feasibility of the DL based environmental awareness in BM for IRS assisted mmWave networks is verified, especially for quasi-static scenarios.

\begin{figure}[tp]
  \centering
  \includegraphics[width=0.45\textwidth]{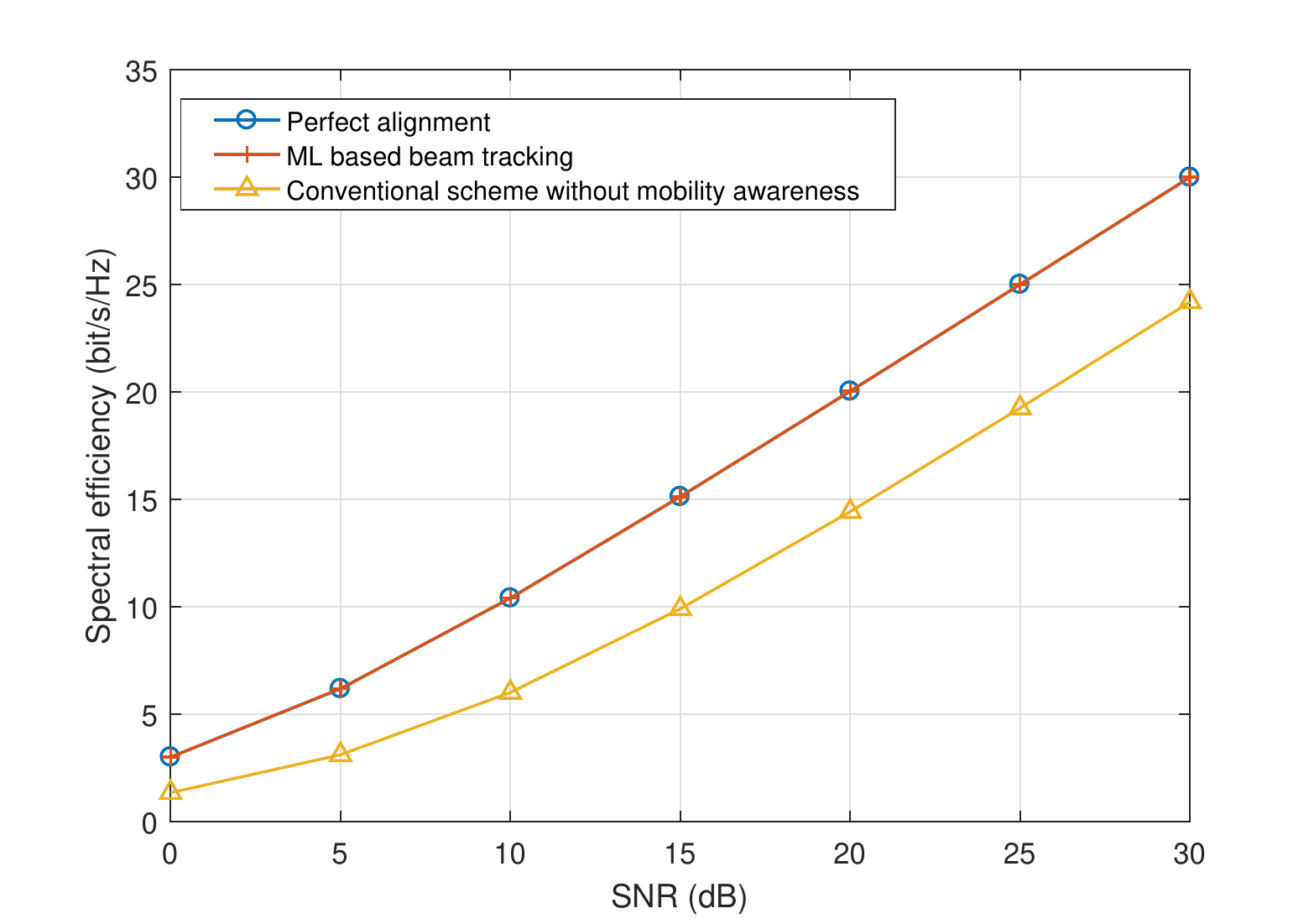}
  \caption{The SE comparison over different beam tracking schemes for high-mobility applications.}
  \label{mobility_awareness}
\end{figure}

For high-mobility scenarios, such as UAV communications, the historical motion state information is utilized to capture the mobility features. For example, we take the dynamic neural network as the mobility prediction model to predict the position of the mobile terminals online. When the predicted position information is input into the environmental-awareness model, and the corresponding optimal IRS selection is further predicted, which facilitates the fast handover for beam-switching among IRSs. Specifically, we compare mobility awareness based beam tracking with conventional scheme without position prediction in terms of the spectral efficiency (SE) performance as shown in Fig. \ref{mobility_awareness}. Notably, our scheme reaps better superiority that its counterpart, and can approach optimal SE performance.

\section{Future Research Directions}

Based on the above explorations for the BM issues and the potential of our proposed ML empowered BM framework, we further suggests some potential research directions for future researches.
\subsection{BM for IRS Assisted Multi-cell Networks}

The current research on BM mainly focuses on conventional single-cell one-hop mmWave networks, while there is few discussion for IRS assisted mmWave networks, specifically for multi-cell networks. On one hand, as mentioned early, the existing BM mechanisms are difficult to straightforwardly extend to the new two-hop communication topology. On the other hand, there is still a hard way to go from single-cell BM to multi-cell BM for IRS assisted mmWave networks. Specifically, the coexist of multiple IRSs and multiple cells makes the network topology more complicated. When it comes to BM, the users need to determine the connected IRS and BS at the same time, unlike the single-cell scenarios where the users only need to select the connected IRS. Therefore, more agile and efficient BM is required to address the challenges for IRS assisted multi-cell networks.

\subsection{ML Empowered BM Framework with Edge Intelligence}

Facing increasingly complicated network topology and diversified service demands for IRS assisted new paradigm, it is very imperative to explore the new ML tools to achieve more efficient and agile BM, which usually requires new \emph{awareness} capabilities, such as service awareness, context awareness and etc.. However, the implementation of such awareness need the support of enough training data and powerful computing power for centralized servers, which will significantly degrade the efficiency of BM. Aiming to overcome the limitations of computing power and training dataset, edge intelligence, also named edge learning, may be a potential solution by exploiting the mobile edge computing platform and the massive distributed data collected by a large number of edge devices. Therefore, the complicated learning tasks can be offloaded to the edge servers, and further be tackled efficiently with a distributed manner.

\subsection{Multi-source Information Fusion Based New BM Framework}

With the development of sensor hardware technology, the mobile terminals can be widely equipped with multiple sensors, such as GPS, Speed, and Visual Sensors and etc.. In particular, vision information is composed of some essential prior information for BM, such as blockage, channel condition, beam direction and etc.. Integrating vision information with the other information obtained by other sensors, a multi-source information fusion based BM framework can be generated to achieve better environmental and mobility awareness, which further enhance BM performance for IRS assisted mmWave networks.

\section{Conclusion}

In this paper, we presents a comprehensive research on BM for IRS assisted mmWave networks. We have analyzed and presented the key issues and challenges of BM for the new network topology, and then the adaptability of existing BM mechanisms that are adaptive to the conventional one-hop mmWave networks is carefully discussed. We have also established a ML empowered BM framework with environment and mobility awareness to address the initial access and frequent-handover challenges of conventional BM in the IRS assisted mmWave networks. A study case is designed to verify the feasibility and potential of our proposed BM framework. As a brief look-forward, we have also outlined some potential research directions to provide sights for future exploration.


%

%

\ifCLASSOPTIONcaptionsoff
  \newpage
\fi



%

\bibliographystyle{IEEEtran}
\bibliography{BM}

\end{document}